\documentclass[12pt]{article}

\usepackage{amssymb, amsmath,hyperref}
\usepackage[utf8]{inputenc}
\usepackage{graphicx}
\usepackage{caption}
\usepackage{cite}
\usepackage{color}
\usepackage{float}

\newcommand{\td}{\text{d}}

\def\be{\begin{equation}}
\def\ee{\end{equation}}
\def\bea{\begin{eqnarray}}
\def\eea{\end{eqnarray}}

\title{ \bf{Soliton mechanics}}

\author{Sharmila Gunasekaran\footnote{sdgg82@mun.ca}, Uzair Hussain\footnote{uh1681@mun.ca}, and Hari K. Kunduri\footnote{hkkunduri@mun.ca }  \\ \\
 \small \sl Department of Mathematics and Statistics, \\  \small \sl  Memorial University of Newfoundland \\ \small \sl St John's NL A1C 4P5, Canada
 }

\date{}

\begin{document}

\maketitle

%\vskip1.5cm

\begin{abstract}
The domain of outer communication of five-dimensional asymptotically flat stationary spacetimes
may possess non-trivial 2-cycles (bubbles). Spacetimes containing such 2-cycles can have non-zero energy, angular momenta, and charge even in the absence of horizons. A mass variation formula has been established for spacetimes containing bubbles and possibly a black hole horizon. This `first law of black hole and soliton mechanics' contains new intensive and extensive quantities associated to each 2-cycle.  We consider examples of such spacetimes for which we explicitly calculate these quantities and show how regularity is essential for the formulae relating them to hold.  We also derive new explicit expressions for the angular momenta and charge for spacetimes containing solitons purely in terms of fluxes supporting the bubbles.   
\end{abstract}

\newpage

%\tableofcontents
%\newpage
%%% Section Introduction
\section{Introduction}
\label{sec:topology}

A striking feature of Einstein-Maxwell theory in four dimensions is the absence of globally stationary, asymptotically flat solutions with non-zero energy - that is, there are `no solitons without horizons' \cite{Gibbons:1997cc}. This property is closely linked to uniqueness theorems for black holes, and indeed it fails to hold in Einstein-Yang Mills theory for which `hairy' black holes exist (see, e.g. \cite{Ashtekar:2000nx}).   In five and higher dimensions, however, non-trivial topology in the spacetime can support the existence of such horizonless solitons even in Einstein-Maxwell supergravity theories.  For an asymptotically flat solution, the topological censorship theorem \cite{Friedman:1993ty} asserts that the domain of outer communication of a spacetime must be simply connected. In four dimensions, that is sufficient to ensure the absence of any cycles in the exterior. In five dimensions, simple connectedness is a weaker constraint, and in particular does not exclude the possibility of 2-cycles (`bubbles'). Physically, these cycles are supported by magnetic flux supplied by Maxwell fields and contribute to both the energy and angular momenta of the spacetime. 

In this note we will focus on five-dimensional asymptotically flat stationary spacetimes with two commuting rotational Killing fields, possibly containing a single black hole.  In this case it has been shown that the topology of the domain of outer communication is $\mathbb{R}\times \Sigma$, where\footnote{In fact, the statement regarding $\Sigma$ is still true if only one rotational Killing field is assumed, although then there are more possibilities for the horizon topology~\cite{Hollands:2010qy}.} 
\be
\Sigma \cong \Big( \mathbb{R}^4 \# n (S^2\times S^2) \# n' (\pm \mathbb{CP}^2) \Big) \backslash B, \label{Sigma}
\ee  
for some $n,n' \in \mathbb{N}_0$ and $B$ is the black hole region, where the horizon $H= \partial B$  must topologically be one of $S^3, S^1 \times S^2 \text{ or } L(p,q)$~\cite{Hollands:2007aj, Hollands:2008fm, Hollands:2010qy, Hollands:2012xy}. The integers $n,n'$ determine the 2-cycle structure of $\Sigma$. 

In the absence of black holes,  soliton spacetimes with 2-cycles supported by flux are known to exist, with a large number of supersymmetric (see the review \cite{Bena:2007kg}) and non-supersymmetric examples \cite{Bena:2009qv,Compere:2009iy, Bobev:2009kn}. These spacetimes carry positive energy.  The relationship between the mass of these spacetimes and their fluxes is expressed in a Smarr-type formula, as observed for BPS solitons in supergravity theories by Gibbons and Warner \cite{Gibbons:2013tqa}. Subsequently, it was shown that under stationary, $U(1)^2$-invariant variations satisfying the linearized field equations, variations of the  mass and magnetic fluxes for general soliton spacetimes are governed by a `first law' formula \cite{Kunduri:2013vka} (see \eqref{mass_var_sol} below).

Furthermore, one can derive a generalised mass and mass variation formula  for $\mathbb{R} \times U(1)^2$-invariant spacetimes containing a black hole with an arbitrary number of 2-cycles in the exterior region. Similar to the soliton case it was found that on top of the familiar terms for a black hole, extra terms due to the bubbles are present. However,  unlike the pure soliton case, these additional terms are most naturally expressed in terms of variations of an intensive quantity (a potential), as opposed to an extensive quantity (a flux). For Einstein-Maxwell theory, possibly with a Chern-Simons term, the mass formula is \cite{Kunduri:2013vka},
\be
 M = \frac{3 \kappa A_H}{16 \pi} + \frac{3}{2} {\Omega}_i  {J}_i + \Phi_H Q+  \frac{1}{2}\sum_{[C]}  {\cal Q}[C]  \Phi[C]+   \frac{1}{2}\sum_{[D]}{\cal Q}[D]   \Phi[D]  \label{BHmass}
 \ee
 and the first law of black hole mechanics is,
 \be
 \delta M = \frac{ \kappa \delta A_H}{8\pi} + \Omega_i \delta J_i + \Phi_H \delta Q + \sum_{[C]} {\cal Q}[C]  \delta  \Phi[C] +\sum_{[D]}  {\cal Q}[D]  \delta \Phi[D]    \; .  \label{BHmech}
\ee In the above $[C]$ is a basis for the second homology of $\Sigma$, $[D]$ are certain disc topology surfaces which extend from the horizon, $\Phi$ are magnetic potentials and  $\cal{Q}$ are certain `electric' fluxes defined on these surfaces which we will define precisely below. This shows that non-trivial spacetime topology plays an important role in black hole thermodynamics, thus providing further motivation to study such objects beyond the obvious implications for black hole non-uniqueness \cite{Kunduri:2014iga}.

It should be noted that most explicitly known examples of soliton spacetimes are supersymmetric, in which case the mass variation formula simply follows from the BPS relation. The same is true for the supersymmetric solution describing a rotating black hole with a soliton in the exterior region \cite{Kunduri:2014iga}. Indeed quite generally for BPS black hole solutions one can show that the additional terms arising in \eqref{BHmass} and \eqref{BHmech} vanish identically.  This is analogous to the fact that for BPS black holes in these theories, the surface gravity and angular velocities also vanish identically.  For non-supersymmetric solutions describing black holes with exterior bubbles, however, these terms would generically contribute.   Examples of such solutions are not explicitly known, although there seems to be no obstruction to their existence, even in the vacuum. 

The purpose of this paper is to apply the formalism developed in  \cite{Kunduri:2013vka} to explicitly compute the various potentials and fluxes appearing above for some known spacetimes with non-trivial $\Sigma$. In so doing we will verify the first variation formula above.  We will also derive some new relations that show how the angular momenta and total electric charge of a spacetime may arise solely from the presence of flux through the 2-cycles. Finally, we will reexamine the singly-rotating dipole black ring \cite{Emparan:2004wy}.  The solution is characterized by a local dipole `charge' resulting from magnetic flux through the $S^2$ of the ring horizon.   The first law for black rings derived in \cite{Copsey:2005se} contains additional terms due to the dipole charge and we show how this is recovered using the general formalism of  \cite{Kunduri:2013vka}.  This will use in a crucial way the disc topology region that lies in the domain of outer communication of the black ring. 
  
\section{First law for black holes and solitons in supergravity}  
The mass and mass variation formulae for asymptotically flat, stationary spacetimes invariant under two commuting rotational symmetries has been established for a general five-dimensional theory of gravity coupled to an arbitrary set of Maxwell fields and uncharged scalars.  We will be concerned with specific soliton and black hole solutions to five-dimensional minimal supergravity, whose bosonic action is (setting Newton's constant $G_5 =1$)
\begin{equation}\label{minsugra}
S = \frac{1}{16\pi} \int_\mathcal{M} \left( \star R -2 F \wedge \star F -\frac{8}{3\sqrt{3}} F \wedge F \wedge A\right)
\end{equation} Here $F = \td A$ and $A$ is a locally defined gauge potential. The existence of a non-trivial second homology $H_2$ implies that $F$ is closed but not exact.  The theory can be recovered from the general theory considered in \cite{Kunduri:2013vka} upon setting $I=1$, $g_{IJ}=2$ and $C_{IJK} = 16/\sqrt{3}$. We will follow this convention throughout when appealing to the construction of potentials and fluxes used in \cite{Kunduri:2013vka}.  The equations of motion are
\begin{equation}\label{eom}
R_{ab} = \frac{4}{3} F_{ac}F_b^{~c} + \frac{1}{3}G_{acd}G_b^{~cd}\;, \qquad  \td \star F + \frac{2}{\sqrt{3}} F \wedge F = 0
\end{equation} where $G =\star F$. The central observation of \cite{Gibbons:2013tqa} was that the non-triviality of the second homology $H_2$ makes it more natural to work with $G$ rather than the gauge potential $A$ which cannot be globally defined. 

Let $\xi$ be the stationary Killing field normalized so that $|\xi|^2 \to -1$ at spatial infinity (in the case of a spacetime containing a black hole, $\xi$ is instead identified with the Killing field which is the null generator of the event horizon).  Using the fact that $F$ is closed and invariant under this action, we have a globally defined potential $\Phi_\xi$ defined by
\begin{equation}
\td \Phi_\xi \equiv i_\xi F 
\end{equation} and the requirement $\Phi_\xi \to 0$ at spatial infinity. From the Maxwell equation one may define a closed two-form
\begin{equation}\label{Theta}
\Theta = 2i_\xi G  - \frac{8}{\sqrt{3}}F \Phi_\xi
\end{equation}  If, in addition to being stationary, the spacetime is invariant under a $U(1)^2$ isometry generated by the Killing fields $m_i = (m_1,m_2)$ (normalized to have $2\pi$-periodic orbits), we also have globally defined magnetic potentials
\begin{equation}
\td \Phi_i = i_{m_i} F
\end{equation} and we also fix the freedom by requiring these vanish at an asymptotically flat end. Together $(\xi, m_i)$ generate an $\mathbb{R} \times U(1)^2$ action acting as isometries on $(\mathcal{M},g,F)$.  Using these potentials one can finally deduce the existence of globally defined potentials $U_i$
\begin{equation}\label{U}
\td U_i = i_{m_i} \Theta + \frac{8}{\sqrt{3}} \td\Phi_i \Phi^H_\xi
\end{equation} which are again fixed by requiring they vanish at the asymptotically flat end.  Here $\Phi^H_\xi$ is the pullback of $\Phi_\xi$ to the horizon if a black hole is present in the spacetime; for a pure soliton spacetime this term is ignored.  The potentials and fluxes defined above can be thought of as functions on a 2d orbit space ${\mathcal{B}} \cong \Sigma / U(1)^2  $ \cite{Hollands:2007aj}. The rank of the matrix $\lambda_{ij} = m_i \cdot m_j $ divides the space into two dimensional interior points, one dimensional boundary segments ($\partial \mathcal{B}$) called rods and zero dimensional points that lie on `corners' where the segments intersect. A black hole is represented by a compact rod $I_H \cong H / U(1)^2 $ where the timelike Killing field goes null. There are two non-compact semi-infinite rods corresponding to the two asymptotic axes of rotation extending out to spatial infinity. The rest of $\partial \mathcal{B}$ contains finite rods $I_i$ where an integer linear combination $v^i m_i, v^i \in \mathbb{Z}$ of the rotational Killing fields vanishes.  This orbit space data thus encodes the action of the isometry group and determines the full spacetime topology up to diffeomorphism \cite{Hollands:2007aj}. In particular finite rods represent two-dimensional submanifolds which may have the topology of either $S^2$, or  a closed disc $D$ if the corresponding rod is adjacent to $I_H$. We will discuss specific examples of spacetimes containing such 2-cycles and discs below. 

 For purely soliton spacetimes (i.e. without black holes), the Smarr formula and mass variation reduce to \cite{Kunduri:2013vka} 
\begin{align}
\label{mass_sol}
M &= \frac{1}{2} \sum_{[C]} \Psi[C] q[C] \\ 
\label{mass_var_sol}
\delta M &=  \sum_{[C]} \Psi[C] \delta q[C].
\end{align}
where
\begin{equation}
\label{mflux}
q[C] = \frac{1}{4 \pi} \int_{C} F \quad \text{and} \quad \Psi[C] = \pi v^i U_i 
\end{equation} represent the magnetic flux and magnetic potential associated to each  element of $[C]$. Note that in \eqref{mass_var_sol} the extensive variable $q[C]$ appears naturally in the first law in contrast to \eqref{BHmech}.  

\noindent
Before discussing specific examples, we would like to present new Smarr-type formulae for the angular momenta and electric charge for purely soliton spacetimes as a sum over fluxes through the 2-cycles. These are useful as they demonstrate how a spacetime can possess such conserved charges in the absence of horizons. 

Firstly, consider the angular momenta $J_i$ associated to the rotational Killing field $m_i$ defined by the Komar integrals
\begin{equation}
J[m_i] = \frac{1}{16\pi} \int_{S^3_\infty} \star \td m_i \;.
\end{equation} The Maxwell equation and Killing property of the $m_i$ imply the existence of two closed (though not necessarily exact) two-forms $\Upsilon_i$ defined by
\begin{equation}
\Upsilon_i \equiv 2  i_{m_i} G - \frac{8}{\sqrt{3}} F \Phi_i \;.
\end{equation} Cartan's formula immediately implies the existence of global potential functions $\chi_{ij}$ satisfying $\td \chi_{ij} = i_{m_i} \Upsilon_j$. Note that we can always choose the integration constant so that $\chi_{ij} =0$ on an interval on which $m_i$ vanishes for fixed $j$.  Now using Stokes' theorem
\begin{equation}
J[m_i] = \frac{1}{8\pi}\int_\Sigma \star \text{Ric}(m_i)= \frac{1}{8\pi} \int_\Sigma \left(-\frac{1}{3}\right)\Upsilon_i \wedge F + \frac{4}{3} \td \star (F \Phi_i)
\end{equation}The final term above may be shown to vanish by converting it to an integral over $S^3_\infty$ where $\Phi_i$ vanishes.  We can evaluate this integral over the orbit space $\mathcal{B}$, giving
\begin{equation}
J[m_i] =  \frac{\pi}{6}\int_{\mathcal{B}} \eta^{jk} \td \chi_{ji} \wedge \td \Phi_k = \frac{\pi}{6} \int_{\mathcal{B}} \td [\eta^{jk}\chi_{ji} \wedge \td \Phi_k]
\end{equation} where $\eta^{ij}$ is the antisymmetric symbol with $\eta^{12} = 1$.  The final term can be converted to a boundary term on $\partial\mathcal{B}$, and using the fact that the potentials vanish on the semi-infinite rods $I_\pm$, we are left with
\begin{equation}
J[m_i] = \frac{\pi}{6} \sum_i \int_{I_i} \eta^{jk} \chi_{ji} \td \Phi_k
\end{equation} This can be further simplified by using the fact that each rod is specified by a pair of integers $v^i$, so that $v^i m_i$ vanishes  By definition $v^i \td \Phi_i =0$ on the rod, so that $\Phi[C] \equiv v^i \Phi_i$ is constant.  By an $SL(2,\mathbb{Z})$ change of basis let us define a new basis $(\hat{m}_1,\hat{m}_2)$ for the $U(1)^2$ generators such that $\hat{m}_1 = v^i m_i$.  The other Killing field $\hat{m}_2$ is non-vanishing on the rod except at the endpoints (these correspond to topologically $S^2$ submanifolds in the spacetime).  Note that in the obvious notation, $\hat{\chi}_{1i}, \hat{\Phi}_1$ are constants on the rod. Using $SL(2,\mathbb{Z})$-invariance, $\eta^{jk} \chi_{ji} \td \Phi_k = \eta^{jk}\hat{\chi}_{ji} \td \hat{\Phi}_k$. Putting the above facts together we arrive at
\begin{equation}
J[m_i] = \frac{1}{3}\sum_{[C]} \chi_i[C] q[C] \label{J_i}
\end{equation} where $q[C]$ are the magnetic fluxes associated to a given cycle $C$ and $\chi_i[C] \equiv -\pi \hat{\chi}_{1i} = -\pi v^j \chi_{ji}$ is a constant associated to each  cycle. It is natural to interpret the $\chi_i[C]$ as \emph{magnetic angular momenta potentials} as they encode how the magnetic flux $q[C]$ contribute to the total angular momenta of the spacetime. 

Now let us turn to an expression for the total electric charge $Q$, defined by
\begin{equation}
Q \equiv \frac{1}{4\pi} \int_{S^3_\infty} \star F = -\frac{1}{2\sqrt{3}\pi} \int_\Sigma F \wedge F 
\end{equation} It may appear counterintuitive that magnetic fluxes contribute to the electric charge , but it should be noted that the Maxwell equation in supergravity is self-sourced.  We now proceed to evaluate this over the boundary of the orbit space.  Using the definition of the magnetic potentials, we have
\begin{equation}
Q = \frac{\pi}{\sqrt{3}}\int_{\mathcal{B}} \eta^{ij} \td \Phi_i \wedge \td \Phi_j = \frac{\pi}{\sqrt{3}} \int_{\partial \mathcal{B}} \eta^{ij} \Phi_i \td \Phi_j \; .
\end{equation}  We can now express this as a sum over the 2-cycles using the argument used above for the angular momenta. The result is
\begin{equation}
Q = -\frac{4\pi}{\sqrt{3}}\sum_{[C]} \Phi[C] q[C]
\end{equation} where $\Phi[C] = v^i \Phi_i$ are constant magnetic potentials associated to each 2-cycle with corresponding rod vector $v^i$.

%%%%%%%%%%%%%%%%%%%%%%%%%%%%%%%%%%%%%%Sec: Single soliton spacetime
\section{Examples}
\subsection{Single soliton spacetime}
Our first example is a charged, non-supersymmetric gravitational soliton with spatial slices $\Sigma \cong \mathbb{R}^4 \# \mathbb{CP}^2$ which was concisely analyzed in \cite{Gibbons:2013tqa} (see also \cite{Compere:2009iy} for a discussion of a generalization which is asymptotically AdS$_5$).  In the following we will use a different parametrization which is convenient for our purposes. The equations of motion \eqref{eom} admit the following local solution,  invariant under an $\mathbb{R} \times SU(2)\times U(1)$ isometry:
\begin{eqnarray}
\td s^2 &=& -\frac{r^2 W(r)}{4 b(r)^2} \td t^2 + \frac{\td r^2}{W(r)} + \frac{r^2}{4}(\sigma_1^2 + \sigma_2^2) + b(r)^2 (\sigma_3 + f(r) \td t)^2 \\
F &=& \frac{\sqrt{3q}}{2} \td \left[\left(\frac{1}{r^2}\right)\left(\frac{j}{2}\sigma_3 - \td t\right)\right]
\end{eqnarray} where $\sigma_i$ are left-invariant one-forms on $SU(2)$:
\begin{eqnarray}
\sigma_1 = -\sin\psi \td \theta + \cos \psi \sin \theta \td \phi \, , \quad \sigma_2 = \cos\psi \td \theta + \sin \psi \sin \theta \td \phi \,, \quad \sigma_3 = \td \psi + \cos\theta \td \phi
\end{eqnarray} 
which satisfy $\td \sigma_i = \tfrac{1}{2}\epsilon_{ijk}\sigma_j \wedge \sigma_k$ and $ \psi \sim \psi + 4\pi$, $\phi \sim \phi +  2\pi$, $\theta \in [0,\pi]$ is required for asymptotic flatness. The functions appearing in the metric are given by
\begin{eqnarray}
W(r) &=& 1 - \frac{2}{r^2}(p -q) + \frac{q^2 + 2 p j^2}{r^4} \qquad f(r) = -\frac{j}{2 b(r)^2}\left(\frac{2p - q}{r^2} - \frac{q^2}{r^4}\right) \\
b(r)^2 &=& \frac{r^2}{4}\left(1 - \frac{j^2 q^2}{r^6} + \frac{2j^2 p }{r^4}\right)
\end{eqnarray} where $p,q,j \in \mathbb{R}$.   We will take $m_i = (\partial_{\hat{\psi}}, \partial_\phi)$, $\hat{\psi} = \psi/2$, to be our basis for the generators of the $U(1)^2$ action with $2\pi$-periodic orbits. 

The parameters $(p,q,j)$ in the above local metric  can be chosen to describe a asymptotically flat, charged rotating black holes. However we may obtain a regular soliton spacetime by requiring that the $S^1$ parameterized by the coordinate $\psi$ degenerates smoothly at some $r = r_0$ in the spacetime, leaving an $S^2$ bolt, or bubble. We therefore require $g_{\psi\psi} = b(r)^2$ vanishes at $r_0$. Regularity of the spacetime metric imposes that $W(r_0) = 0$. The existence of a simultaneous root fixes
\begin{equation}\label{cond}
p = \frac{r_0^4(r_0^2 - j^2)}{2 j^4} \qquad q = -\frac{r_0^4}{j^2}
\end{equation} In order for $\partial_{\hat{\psi}}$ to degenerate smoothly and avoid a conical singularity at $r =r_0$ requires $W'(r_0)(b^2(r_0))'=1$, or equivalently
\begin{equation}
(1-x)(2+x)^2 = 1
\end{equation} for $x =x_* =  r_0^2/j^2$.  This cubic has a unique positive solution at $x \approx 0.870385$, and in particular $r_0^2 < j^2$.  

With this inequality it is easy to check that $W(r), b(r)^2>0$ for $r>r_0$ and the spacetime metric is globally regular. Further
\begin{equation}
g^{tt} = - \frac{4 b(r)^2}{r^2 W(r)} < 0
\end{equation}  so the spacetime is stably causal, and in particular the $t=$constant hpersurfaces are Cauchy surfaces. It can be verified that $g_{tt} < 0$ everywhere, so $\partial/ \partial t$ is globally timelike and in particular there are no ergoregions. However, if one uplifts the soliton to six dimensions, we expect it  will suffer from the instability discussed in \cite{Cardoso:2005gj}. 

 We thus obtain a 1-parameter family of $\mathbb{R} \times SU(2) \times U(1)$-invariant soliton spacetime. 

The $S^2$ at $r = r_0$ has a round metric 
\begin{equation}
\td s^2_2 = \frac{r_0^2}{4} (\td \theta^2 + \sin^2\theta \td \phi^2)
\end{equation} and carries a magnetic flux
\begin{equation}
q[C] = \frac{1}{4\pi} \int_{S^2} F = \frac{\sqrt{3} r_0^2}{4 j}
\end{equation}
It is straightforward to read off
\begin{equation}
\Phi_\xi = \frac{\sqrt{3}q}{2 r^2} \;,\qquad \Phi_{\hat{\psi}} =-\frac{\sqrt{3}q j}{2r^2} \;, \qquad \Phi_{\phi} = -\frac{\sqrt{3} q j \cos\theta}{4 r^2}\;.
\end{equation} A long but straightforward calculation yields, using \eqref{Theta} and \eqref{U}:
\begin{align}
\td U_{\hat\psi} &= \left[\frac{2\sqrt{3} j q}{r^3} - \frac{4 \sqrt{3} j q^2}{r^5} \right]\td r \\
\td U_\phi &= \left[ -\frac{2\sqrt{3} j q^2 \cos \theta}{ r^5} + \frac{\sqrt{3} j q \cos \theta}{ r^3} \right] \td r  + \left[ -\frac{\sqrt{3} j q^2 \sin \theta}{2 r^4} + \frac{\sqrt{3} j q \sin \theta}{2 r^2} \right] \td \theta
\end{align}
which leads to 
\begin{align}
U_{\hat \psi} = \frac{\sqrt{3} j q}{r^2}\left(\frac{q}{r^2} - 1\right) \;,\qquad
U_{ \phi} = \frac{\sqrt{3} j q \cos \theta}{ 2r^2} \left( \frac{  q }{ r^2} - 1 \right)
\end{align}
where the integration constants have been fixed so that the potentials vanish as $r \to \infty$. 

On the $S^2$ `bolt' at $r=r_0$, the Killing field $\partial_{\hat\psi} = 2\partial_\psi$ degenerates smoothly.  The interval structure of the orbit space is given below in the basis of rotational Killing fields orthogonal at infinity $(\partial_{\phi_1},\partial_{\phi_2})$ where $\partial _{\phi_1} = \partial_\psi - \partial_\phi $ and $\partial _{\phi_2} = \partial_\phi + \partial_\psi  $. In this basis the two semi-infinite rods can be manifestly seen as axes of rotation with vanishing $\partial_{\phi_1} $ or $\partial_{\phi_2}$.
\begin{figure}[H]
\begin{center}
\includegraphics[scale=0.5]{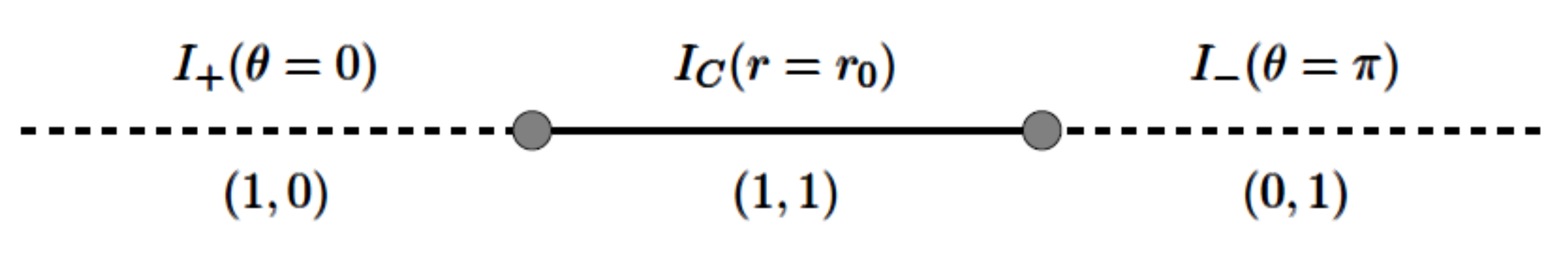}
\captionsetup{format=hang}
\caption{ Rod structure for single soliton spacetime in $(\phi_1, \phi_2)$ basis.}
\end{center}
\end{figure}
 \noindent
 We now turn to the computation of the potentials associated to the soliton. Firstly,
\begin{equation}
\Psi[C] = \pi U_{\hat\psi}(r_0) = \frac{\sqrt{3} \pi r_0^2(j^2 + r_0^2)}{j^3}
\end{equation} We then find
\begin{equation}
\frac{\Psi[C] q[C]}{2} = \frac{3\pi}{8} \left(\frac{r_0}{j}\right)^4 (j^2 + r_0^2) 
\end{equation}  which is indeed the ADM mass of the spacetime, which can easily be read off from the expansion
\begin{equation}
g_{tt} = -1 + \frac{8M}{3 \pi r^2} + O(r^{-4})
\end{equation} Finally the first law of soliton mechanics asserts that
\begin{equation}
\td M = \Psi[C] \td q [C]
\end{equation} In our explicit example, 
\begin{equation}
\td M - \Psi[C] \td q[C] = \frac{3\pi r_0^5}{4j^5}(j \td r_0 - r_0 \td j)
\end{equation} and the right hand side vanishes as a consequence of the regularity condition $r_0^2/j^2 = x_*$. We emphasize that the Smarr-type relation for the mass does not require regularity of the spacetime to hold, whereas the first law is in fact a finer probe of regularity.  Finally one can explicitly check that the electric charge is indeed given by
\begin{equation}\label{charge}
Q = -\frac{4\pi}{\sqrt{3}} \Phi[C] q[C] = -\frac{\sqrt{3} \pi r_0^4}{2 j^2}\;.
\end{equation} 

To compute the magnetic angular momentum potentials $\chi_{ij}$, it is convenient to work in the $U(1)^2$ basis $(\partial_\psi, \partial_\phi)$ and then convert to the basis $(\partial_{\phi_1}, \partial_{\phi_2})$ which is orthogonal at the asymptotically flat end, in order to fix integration constants. A long but straightforward calculation yields
\begin{eqnarray}
\chi_{\psi \psi} & = &-\frac{\sqrt{3} q^2 j^2}{4r^4} + \frac{\sqrt{3}q}{4}\;, \qquad \chi_{\phi\psi} = \frac{\sqrt{3} q \cos\theta}{4}\left(1 - \frac{q  j^2}{r^4}\right) \\
\chi_{\phi \phi}&=& -\frac{\sqrt{3} q^2 j^2 \cos^2 \theta}{4r^4} -\frac{\sqrt{3}q}{4}\;, \qquad  \chi_{\psi \phi} = -\frac{\sqrt{3} q \cos\theta}{4}\left(1 + \frac{q j^2}{r^4}\right) \nonumber
\end{eqnarray} Since the 2-cycle is specified by the vanishing of $\partial_{\hat \psi}$, using the formula \eqref{J_i} we find
\begin{equation}
J_\psi = \frac{\pi r_0^6}{4j^3}\;, \qquad J_\phi = 0
\end{equation} where in the second equality we observe that $\chi_{\psi \phi} =0 $ on $C$ using \eqref{cond} . It is easy to check that these expressions agree with the standard ADM angular momenta computed from the asymptotic fall-off of the metric. As expected, the $SU(2) \times U(1)$-invariant solution has equal angular momenta in orthogonal 2-planes, $J_1 = J_2 = J_\psi$. Note that $J_\psi \neq 0$ for the soliton; indeed, we have the constraint
\begin{equation}
J_\psi = -\frac{2 Q q[C]}{3} = \frac{16\pi q[C]^3}{3\sqrt{3}} \;.
\end{equation}

\subsection{Double soliton spacetime}
Our second example is a supersymmetric, asymptotically flat spacetime containing two non-homologous two-cycles. The spatial slices $\Sigma \cong \mathbb{R}^4 \# (S^2 \times S^2)$ where the connected sum with $\mathbb{R}^4$ corresponds to removing a point.  The solution is originally given in the more general $U(1)^3$ five-dimensional supergravity \cite{Bena:2005va}. We will quickly review the solution to the minimal supergravity theory \eqref{minsugra} as this does not seem to be reproduced explicitly in the literature.  The spacetime metric takes the canonical form of a timelike fibration over a hyperK\"ahler `base space'
\begin{equation}
\td s^2 = -f^2( \td t + \omega)^2 + f^{-1} \td s^2_B \; ,
\end{equation} 
where $V = \partial /\partial t$ is the supersymmetric, timelike Killing vector field and $\td s^2_M$ is a hyperK\"ahler base~\cite{Gauntlett:2002nw}.  The solution has a Gibbons-Hawking hyperK\"ahler base
\begin{equation}
\td s^2_M = H^{-1} (\td \psi + \chi)^2 + H (\td r^2 + r^2 (\td\theta^2 + \sin^2\theta \td \phi^2)) \; ,
\end{equation} 
where $(r,\theta,\phi)$ are spherical coordinates on $\mathbb{R}^3$, the function $H$ is harmonic on $\mathbb{R}^3$ and $\chi$ is a 1-form on $\mathbb{R}^3$ satisfying $\star_3 \td  \chi = \td H$. 

The analysis of~\cite{Gauntlett:2002nw} shows a general technique for constructing solutions of the above form.  Defining the following harmonic functions on $\mathbb{R}^3$~\cite{Bena:2005va} 
\begin{eqnarray}
H &=& \frac{1}{r}  - \frac{1}{r_1}+ \frac{1}{r_2}, \qquad  \qquad K = \frac{k_0}{r} + \frac{k_1}{r_1} + \frac{k_2}{r_2} ,\\
L &=& 1+ \frac{\ell_0}{r}  + \frac{\ell_1}{r_1} + \frac{\ell_2}{r_2}, \qquad  \qquad M = m   + \frac{m_1}{r_1}+ \frac{m_2}{r_2},
\end{eqnarray} with
\be
r_1 = \sqrt{r^2 + a_1^2 - 2ra_{1} \cos\theta}, \qquad \qquad r_2 = \sqrt{r^2 + a_2^2 - 2ra_{2} \cos\theta}   \label{centres}
\ee where we assume $0 < a_1 < a_2$, we arrive at a solution provided
\begin{equation}
 f^{-1} = H^{-1} K^2 + L \; ,\qquad  \omega = \omega_\psi(\td \psi + \chi ) + \hat{\omega} \; ,   \label{f}
 \ee
 where 
 \begin{eqnarray}
 \label{omega}
&& \omega_\psi = H^{-2} K^3 + \frac{3}{2}H^{-1}KL + M \; , \\  && \star_3 \td \hat\omega = H \td M - M \td H + \frac{3}{2}(K \td L - L \td K)
   \; .
 \end{eqnarray} 
 The Maxwell field is then
 \be
 \label{max}
 F = \frac{\sqrt{3}}{2} \td \left[ f( \td t + \omega) - K H^{-1} (\td \psi+ \chi_i \td x^i)  - \xi_i \td x^i\right]  \; ,
 \ee
where the 1-form $\xi$ satisfies  $\star_3 \td \xi = - \td K$.  For the above choice of harmonic functions one finds 
 \be
 \chi =\left[  \cos \theta - \frac{r\cos \theta -a_1 }{r_1}+ \frac{r \cos\theta - a_2}{r_2}  \right] \td\phi \; ,
 \ee  and
 \be
\xi  = -\left[  k_0 \cos \theta + \frac{k_1(r\cos \theta -a_1) }{r_1}+ \frac{k_2(r \cos\theta - a_2)}{r_2}  \right] \td\phi  \; ,
 \ee  where we have absorbed the integration constant in $\chi$ by suitably shifting $\psi$.  One may also integrate explicitly for $\hat\omega = \hat{\omega}_\phi \td \phi$. 

 For a suitable choice of constants this solution is asymptotically flat provided $\Delta \psi = 4\pi$, $\Delta \phi = 2\pi$ and $0 \leq \theta \leq \pi$. In particular setting $r = \rho^2/4$ and sending $\rho \to \infty$ one finds
 \bea
 &&\td s^2_M \sim \td\rho^2 + \frac{\rho^2}{4}\left[(\td\psi+ \cos \theta \td \phi)^2)^2 \td \theta^2 + \sin^2 \theta \td\phi^2\right]
\eea with  $O(\rho^{-2})$ corrections in the associated Cartesian chart. Finally, choosing
\begin{equation}
 m = -\frac{3}{2}(k_0 + k_1 +k_2)   \label{m}
 \end{equation} and suitably fixing the integration constant in $\hat\omega_\phi$, we find $f = 1 + \mathcal{O}(\rho^{-2})$, $\omega_\psi= \mathcal{O}(\rho^{-2})$ and $\hat{\omega}_\phi = \mathcal{O}(\rho^{-2})$ . Thus the spacetime is asymptotically Minkowski $\mathbb{R}^{1,4}$. 

The free parameters characterizing these local `three-centre' solutions may be chosen so that globally, the spacetime describes a two-soliton spacetime (see, e.g. \cite{Gibbons:2013tqa}).  It is clear that the spacetime metric is regular apart from possible singularities at the `centres' which lie at the points ${\bf x_0} = (0,0,0)$,  ${\bf x_1} = (0,0,a_1)$, and ${\bf x_2} = (0,0,a_2)$ in the usual Cartesian coordinates on the ambient $\mathbb{R}^3$ on the base space. To ensure that the spacetime metric degenerates smoothly at these points, it is sufficient to first require that the base space be smooth. It can be shown that this is in fact the case without any further restriction of parameters (the base space metric approaches, up to a overall sign, the Euclidean metric near the origin of $\mathbb{R}^4$). Note that on the base space, $\partial_\psi$ degenerates smoothly at the centres.  

Next to ensure that the spacetime metric is well behaved and has the correct signature, we must have $f\neq 0$ ($f=0$ would correspond to an event horizon).  Equivalently we must ensure $f^{-1}$ does not diverge, which fixes
\begin{eqnarray}
\ell_2 = -k_2^2 \,, \quad \ell_1 = k_1^2 \,, \quad \ell_0 = -k_0^2 \;.
\end{eqnarray} Further, since $\partial_\psi$ degenerates on the base, near the centres we have \begin{equation}
|\partial_\psi|^2 = -f^2 \omega_\psi^2 \leq 0
\end{equation} which immediately implies that $\omega_\psi$ must \emph{vanish} at these points.  It turns out generically $\omega_\psi$ actually has simple poles at these points. Removing these requires
\begin{equation}
m_1 = \frac{k_1^3}{2} \,, \quad m_2 = \frac{k_2^3}{2} \,, \quad k_0 = 0\,.
\end{equation} Actually imposing that $\omega_\psi=0$ leads to the so called `bubble equations'
\begin{gather}\label{constraints}
a_2 k_1^3 + a_1 k_2^3 - 3a_1 a_2(k_1 + k_2) = 0\\
a_1(k_1 + k_2)^3 + (a_2 - a_1)(k_1^3 -3a_1(2k_1 +k_2)) = 0 \\
a_2(k_1 + k_2)^3 -(a_2 - a_1)(k_2^3 + 3a_2 k_1) = 0
\end{gather} which correspond to the enforcing regularity at $r=0, r=a_1$, and $r=a_2$ respectively.  This leaves a one-parameter family of 2-soliton spacetimes parameterized by $(a_1,a_2, k_1,k_2)$ subject to the three regularity constraints.  An analysis of the geometry shows that the spacetime is stably causal ($g^{tt} \leq 0$) \cite{Gibbons:2013tqa}. 

Let us now consider  the boundary structure of the orbit space $\mathcal{B} = \Sigma/U(1)^2$, which determines the topology of the spacetime.  There is a semi-infinite rod $I_+$ corresponding to one of axes of symmetry in the asymptotically flat region. The appropriately normalized Killing field which vanishes on this rod is $v_+ = \partial_\psi - \partial_\phi$. In terms of the spherical coordinates on the ambient $\mathbb{R}^3$ associated to the Gibbons-Hawking space, $I_+ =\{ r>a_2, \theta =0\}$. Next, there is a finite rod $I_{C_2} = \{a_1 < r < a_2, \theta =0\}$  with associated vanishing Killing field $v_2 = -(\partial_\phi + \partial_\psi)$. Note that the Killing field $\partial_\psi$ is non vanishing on $C_2$ and degenerates smoothly at the endpoints $r =a_1, a_2$ implying that $C_2$ is a topologically $S^2$-submanifold in the spacetime. The second `bubble' corresponds to the interval $I_{C_1} = \{0 < r < a_1, \theta =0\}$ with associated Killing field $v_1 = -\partial_\phi + \partial_\psi$.  The Killing field $\partial_\psi$ is again non-vanishing on this interval and degenerates smoothly at the endpoints $r=0, r=a_1$.  Finally, there is a second semi-infinite rod $I_- = \{ r>0,\theta = \pi\}$ with associated Killing field $v_- = \partial_\phi + \partial_\psi$. 

The rod structure is most naturally expressed in terms of the basis of Killing fields $m_1 = v_+, m_2 = v_-$ which have $2\pi$ periodic orbits: 
\begin{equation}
v_+ = (1,0) \,, \quad v_2 = (0,-1)\,, \quad v_1 = (1,0) \, ,\quad v_- = (0,1) \,
\end{equation} from which it is easy to check that the compatibility condition $|\det (v_i^T v_{i+1}^T)| = 1$ is satisfied for adjacent rods. 
\begin{figure}[H]
\begin{center}
\includegraphics[scale=0.5]{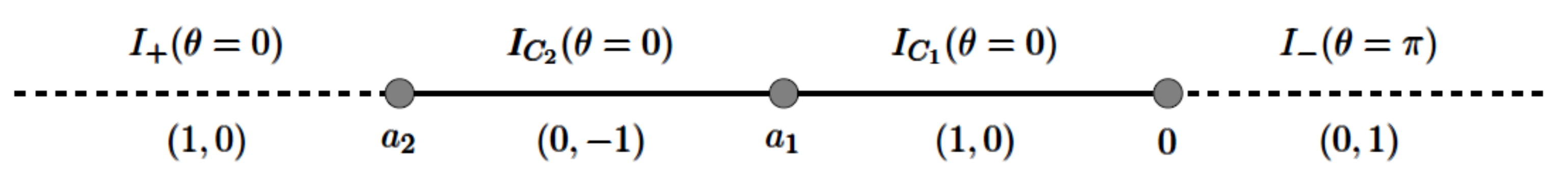}
\captionsetup{format=hang}
\caption{ Rod structure for double soliton spacetime in $(\phi_1, \phi_2)$ basis. Here, $\partial _{\phi_1} = \partial_\psi - \partial_\phi $ and $\partial _{\phi_2} = \partial_\phi + \partial_\psi  $. }
\end{center}
\end{figure}
\noindent
We now turn to a computation of the various intensive and extensive quantities appearing in the first law. The magnetic fluxes through the bubbles $C_1, C_2$ are found to be
\begin{equation}\label{q[C]}
q[C_2] = \frac{1}{4\pi}\int_{S^2_{2}} F = -\frac{\sqrt{3}}{2} (k_1 + k_2) \,,\qquad q[C_1] = \frac{1}{4\pi} \int_{S^2_{1}} F = \frac{\sqrt{3}}{2} k_1
\end{equation} The computation of the `electric' potentials $U_i$ requires some more work.  For a general supersymmetric solution in the timelike class, one can derive the relation
\begin{equation}
i_\xi \star F = \frac{\sqrt{3}}{2} f^2 \star_4 \td \omega - \frac{f G^+}{\sqrt{3}}
\end{equation} where $\star_4$ is the Hodge dual taken with respect to the base space and $G^+ = \tfrac{f}{2}(\td \omega + \star_4 \td \omega)$ is a self-dual 2 form.   Using this, and the general form of the Maxwell field leads to the simple expression
\begin{equation}
\Theta = \sqrt{3} \td (f^2(\td t + \omega)) - 4F
\end{equation} from which it is manifest that $\Theta$ is closed, though not exact, as expected.  We then have
\begin{align}
U_\psi &= -\sqrt{3} f^2 \omega_\psi + 4 A_\psi + 2\sqrt{3}(k_1 + k_2)\\
U_\phi &= -\sqrt{3}f^2 \omega_\phi + 4 A_\phi
\end{align} where $A_\psi, A_\phi$ are the components of the gauge field and  integration constants have been chosen so that $U_i$ vanish at spatial infinity. As discussed above, $v^i_{C_2} U_i$ and $v^i_{C_1} U_i$ must be constant on the two-cycles $C_2 $ and $C_1$ respectively.  In order to demonstrate this, one must make use of the regularity constraints \eqref{constraints}.  We find
\begin{gather}
\Psi[C_2]  = \pi U_{C_2} \equiv -\pi (U_\psi + U_\phi)\vert_{I_{C_2}} = -4\sqrt{3}k_1   \\  \Psi[C_1] = \pi U_{C_1} \equiv \pi (U_\psi - U_\phi)\vert_{I_{C_1}} = 4\pi \sqrt{3}(k_1 + k_2) 
\end{gather} Using this we can indeed verify that 
\begin{equation}\label{smarr2soliton}
\frac{1}{2}\sum_C \Psi[C] q[C]  = 6\pi k_1 (k_1 + k_2) = M
\end{equation} The first law 
\begin{equation}\label{BPSfirst}
\delta M = \Psi[C_1] \delta q[C_1] + \Psi[C_2] \delta q[C_2]
\end{equation} can then be verified explicitly (we emphasize this is independent from \eqref{smarr2soliton}).  Note that it is straightforward to check that the magnetic potentials are 
\begin{equation}
\Phi[C_1] =-\sqrt{3}(k_1 + k_2) =  -\frac{1}{4\pi} \Psi[C_1] \;,\qquad \Phi[C_2] =\sqrt{3}k_1 =  -\frac{1}{4\pi} \Psi[C_2]
\end{equation} and inserting these into \eqref{charge} for the total electric charge expressed as sum over the basis of 2-cycles,  one recovers the usual BPS relation $M =\sqrt{3} Q/2$.  The variational formula \eqref{BPSfirst} is surprising as it represents a genuine `first law' for BPS geometries, whereas for BPS black holes, the first law trivially follows from the BPS condition (i.e. $\delta M = \sqrt{3}\delta Q/2$).

The calculation of angular momenta from the general formula \eqref{J_i} is less straightforward. The difficulty arises from the complexity of the solution, and although it is possible to show that $\td \chi_{ij} =0$, obtaining the integrated potentials in closed form has proved difficult.  However, it should be noted that the asymptotic conditions $v_+^i \chi_{ij} =0$ on $I_+$ and $v_{-}^i \chi_{ij}$ on $I_-$, as well as the evaluation of $\chi_{i}[C]$ on each cycle, only require knowledge of $\chi_{ij}$ on the `axes' $\theta =0,\pi$.  Hence we need only integrate for $\chi_{ij}(r,0)$ and $\chi_{ij}(r,\pi)$ on each segment on the axis (i.e. $I_\pm, I_{C_i}$).  Since the $\chi_{ij}$ must be continuous functions of $r$ along the axes across the rod points at $r=a_2, r=a_1$, and $r=0$, the integration constants arising from integrating separately over each segment  are determined completely by the asymptotic conditions.  Carrying this out carefully one finds
\begin{equation}
\chi_\phi[C_2] = 2\sqrt{3}k_1(k_1 + 2k_2)\;, \qquad  \chi_\phi[C_1] = -2\sqrt{3}(k_2^2 - k_1^2)
\end{equation} and
\begin{equation}
\chi_\psi[C_2] =  -2\sqrt{3}k_1(3k_1 + 2k_2)\;, \qquad \chi_\psi[C_1]  = 2\sqrt{3}(3k_1^2 + 4k_1 k_2 + k_2^2)
\end{equation} where we have used the regularity constraints \eqref{constraints} to significantly simplify these expressions.  Using the expressions for the fluxes \eqref{q[C]} we obtain the angular momenta
\begin{equation}
J_\psi = 3\pi k_1(k_1 + k_2)(2k_1 + k_2) \;, \qquad J_\phi =-3\pi k_1 k_2(k_1 + k_2) \;,
\end{equation} which do in fact agree with the standard ADM angular momenta provided that \eqref{constraints} is used to simplify the latter.  

Using the above expressions for the charges $(J_\psi, J_\phi, Q)$ and fluxes $q[C_i]$, we can derive
\begin{eqnarray}
J_\psi &=& =\frac{Q}{2} (q[C_1] - q[C_2]) = \frac{8\pi}{\sqrt{3}} q[C_1] q[C_2]\left(q[C_2] - q[C_1]\right)\;, \\
J_\phi &=&  \frac{Q}{2} (q[C_2] + q[C_1]) =  -\frac{8\pi}{\sqrt{3}} q[C_1] q[C_2]\left(q[C_2] + q[C_1]\right) \;.
\end{eqnarray} The angular momenta about the $\psi-$ and $\phi-$ directions thus is a measure of the difference and sum of the magnetic fluxes out of the two bubbles.

%%%%%%%%%%%%%%%%%%%%%%%%%%%%% - Section : dipole ring
\subsection{Dipole black ring}

As a last example, we consider asymptotically flat dipole black rings\cite{Emparan:2004wy} where the horizon topology is $ S^1 \times S^2$ and $\Sigma \cong \mathbb{R}^4 \# (S^2 \times D^2)$ \cite{Alaee:2013oja,Andersson:2015sfa} .The rings are a solution to five dimensional Einstein-Maxwell theory (and also the minimal supergravity theory because the Chern-Simons term is of no consequence to the solutions). For convenience to match with the conventions used in \cite{Emparan:2004wy}, in this section we take $g_{IJ} = 1/2$ in the general formalism of \cite{Kunduri:2013vka}. The metric is given by
\begin{align} \label{dipolering}
ds^2= & -\frac{F(y)}{F(x)}\left(\frac{H(x)}{H(y)}\right)
\left(\td t+C(\nu,\lambda)\: R\:\frac{1+y}{F(y)}\: \td\psi\right)^2\\[3mm]
&+\frac{R^2}{(x-y)^2}\: F(x)\left(H(x)H(y)^2\right)\left[
-\frac{G(y)}{F(y)H(y)^3}\td \psi^2-\frac{\td y^2}{G(y)}
+\frac{\td x^2}{G(x)}+\frac{G(x)}{F(x)H(x)^3}\td\varphi^2\right] \nonumber
\end{align}
with the gauge potential,
\begin{equation}
A_\varphi = \sqrt{3} C(\nu, -\mu) R \frac{1+x}{H(x)}
\label{ringvec} 
\end{equation}
The functions in the metric are defined as follows, 
\begin{align}
F(\xi) & = 1 + \lambda \xi,  \quad G(\xi) = (1 - \xi^2)(1 + \nu \xi), \quad H(\xi) = 1 - \mu \xi \\  \text{ with } & 0 < \nu \leq  \lambda < 1\,, 0 \leq \mu < 1 \text{ and } C(\alpha, \beta) = \sqrt{ \beta(\beta - \alpha) \frac{1 + \beta}{ 1 - \beta}}, \nonumber  
\end{align}
where $\alpha$ and $\beta$ are any two of the parameters $\mu, \nu \text{ and } \lambda$. 

\noindent
The following relations remove conical singularities at $y=-1$, $x=- 1$ and $x = +1$. 
\begin{equation}\label{period0}
\Delta\psi=\Delta\varphi =
2\pi\frac{(1+\mu)^{3/2}\sqrt{1-\lambda}}{1-\nu}\,, \quad  \frac{1-\lambda}{1+\lambda}\left(\frac{1+\mu}{1-\mu}\right)^3= \left(\frac{1-\nu}{1+\nu}\right)^2\,
\end{equation}
Thermodynamic quantities for \eqref{dipolering} were calculated in \cite{Emparan:2004wy}. Here, we specifically focus on rederiving the the extra terms that contribute to the mass using the results in \cite{Kunduri:2013vka}. These extra terms arise from disc topology surfaces denoted by $D$  that meet the horizon. The fluxes and potentials evaluated on these surfaces can be done so on any other surface that is homologous to $D$ with the same boundary as $D$. Studying the rod structure of the solution reveals a disc topology surface at $x=1$.
\begin{figure}[H]
\begin{center}
\includegraphics[scale=0.5]{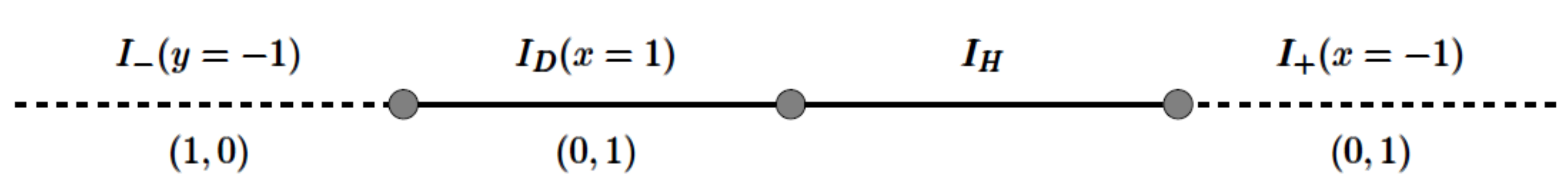}
\captionsetup{format=hang}
\caption{ Rod structure for dipole ring}
\end{center}
\end{figure}
\noindent
 The disc $D$ is parametrized by $(y, \psi) $ at constant $ t $, $\phi$ and $x = 1$. 
The flux ${\cal{Q}}[D]$ is given by 
\begin{equation}
{\cal{Q}}[D] = \int_{[D]} \Theta  = -\frac{\sqrt{3} \pi  (\mu +1) R \sqrt{ \mu (1 -\lambda) (1 - \mu)   }}{4 \sqrt{(\mu +\nu )}}
\end{equation}
(For usual Einstein-Maxwell theory $g_{IJ} = \frac{1}{2}$ and $C_{IJK} = 0$). $\partial_\varphi $ vanishes at $x=1$.
$(v^1,v^2) = (0,1)$ in the $(\hat{\partial_\psi},\hat{\partial_\varphi})$ basis, where the Killing fields are normalized to have $2 \pi$ periodic orbits. 
\begin{equation}
\Phi[D] = v^i \Phi_i = -\frac{2 \sqrt{3} (1+\mu) R \sqrt{ \mu (1 - \lambda )   (\mu +\nu )}}{\sqrt{(1 -\mu)} (1 - \nu)}
\end{equation}

It is easily checked that the potential $\Phi[D] = -2\cal{D}$ and flux $\mathcal{Q}[D]= -\frac{1}{2}\hat{\Phi}$  where $\cal{D}$ is the local dipole charge and $\hat{\Phi}$ is the magnetic potential introduced\footnote{The quantities $\cal{D}$ and $\hat{\Phi}$ are referred to as  $\cal{Q}$ and $\Phi$ respectively in the notation of \cite{Emparan:2004wy}. We are using different symbols to avoid confusion with the notation of \cite{Kunduri:2013vka}.} in \cite{Emparan:2004wy}. Therefore, we see that the Smarr relation and first law given in \cite{Emparan:2004wy} 
\begin{equation}
M = \frac{3}{16\pi} \kappa A_H+ \frac{3}{2} \Omega_H J + \frac{1}{2} \mathcal{D}\hat{\Phi} \;, \qquad \delta M = \frac{\kappa \delta A_H}{8\pi} + \Omega_H \delta J + \hat{\Phi} \delta \mathcal{D}
\end{equation} match precisely with the derived expressions in \eqref{BHmass} and \eqref{BHmech}. An important point to emphasize is that, although the local dipole charge $\mathcal{D}$ arises as a \emph{flux} integral of $F$ over the $S^2$ of the black ring \cite{Emparan:2004wy}, in our formalism  it arises as the constant value of $\Phi$ evaluated on the equipotential disc surface $D$ which ends on  the horizon. Hence, although it seems counterintuitive that variations of an `intensive' variable such as $\Phi[D]$ appear in the general first law, we see that at least in the present case, it is more naturally interpreted as an extensive variable (the dipole charge). Indeed if one looks at the fall-off of the gauge field $A$ at the asymptotically flat region \cite{Emparan:2006mm}, this quantity can be interpreted as producing a dipole contribution.  The fact that $\Phi[D]$ captures, in an invariant way, the dipole charge has also been observed in the context of black lenses \cite{Kunduri:2014kja,  Kunduri:2016xbo, Tomizawa:2016kjh}. In the case of black lenses, there is in fact no natural 2-cycle in the spacetime on which to define a dipole chrarge as there is for a ring \cite{Kunduri:2016xbo}. 
\section{Discussion} 
We have explicitly computed the additional terms in the Smarr relation and first law arising from non-trivial spacetime topology in three different geometries, two describing solitons and another describing a black ring. For purely soliton spacetimes, we have complemented the results in \cite{Kunduri:2013vka} with a Smarr type formula for $J$ and $Q$. These expressions also demonstrate the presence of conserved charges in the absence of a horizon. We have seen that spacetime regularity is crucial for the first law to be satisfied for all examples.
 
A conjectured relation \cite{Gubser:2000mm} between dynamical and thermodynamic instability has been established by Hollands and Wald \cite{Hollands:2012sf}. They have shown that the black p-brane spacetime $M \times \mathbb{T}^p$ associated to a thermodynamically unstable black hole $M$ is itself dynamically unstable.  This result of course applies to spacetimes with horizons only, and do not pertain to the soliton spacetimes considered here. Very recently, the linear stability of supersymmetric soliton geometries has been investigated \cite{Eperon:2016cdd} (see also \cite{Keir:2016azt} for a rigorous analysis of the scalar wave equation).  In particular the authors of \cite{Eperon:2016cdd} have produced evidence that these solutions suffer from a non-linear instability associated with the slow decay of linear waves.  It would be interesting if a  connection could be found between these studies of dynamical instability and an analogue of thermodynamic instability using the laws of soliton mechanics discussed in this work. \\

\noindent  {\bf Acknowledgements} \\

\noindent HKK is supported by an NSERC Discovery Grant. This research was supported in part by Perimeter Institute for Theoretical Physics. Research at Perimeter Institute is supported by the Government of Canada and by the Province of Ontario. We thank James Lucietti for a number of useful suggestions and for comments on the draft.

\bibliography{soliton}
\bibliographystyle{utphys}
\end{document}